\documentclass[aoas,MSNbibl,nameyear,seceqn,dvips]{arximspdf}
\usepackage{dcolumn}
\usepackage{graphicx}

\doi{10.1214/12-AOAS601} 
\volume{7}
\issue{1}
\pubyear{2013}
\firstpage{570}
\lastpage{587}

\makeatletter
\newcommand{\rrvert}{\vert}
\newcommand{\llvert}{\vert}
\newcolumntype{d}[1]{D{.}{.}{#1}}
\def\cal{\mathcal}

\newcommand{\cov}{\operatorname{Cov}}
\newcommand{\cor}{\operatorname{Corr}}
\newcommand{\var}{\operatorname{Var}}
\newcommand{\E}{\mathrm{E}}
\newcommand{\diag}{\operatorname{diag}}
\newcommand{\Zj}{\mathbf{Z}_j}
\newcommand{\muZ}{\bolds{\mu}_{Z}}
\newcommand{\Bbeta}{\bolds{\beta}}
\newcommand{\GammaZ}{\bolds{\Gamma}_{Z}}
\newcommand{\GammagZ}{\bolds{\Gamma}_{f Z}}
\newcommand{\BGamma}{\bolds{\Gamma}}
\newcommand{\g}{\mathbf{f}}
\newcommand{\Ba}{\mathbf{a}}
\newcommand{\BB}{\mathbf{B}}
\makeatother

\begin{document}
\begin{frontmatter}

\title{Canonical correlation analysis between time series and static
outcomes, with application to the spectral analysis of heart rate variability}
\runtitle{Time series CCA}

\begin{aug}
\author[a]{\fnms{Robert T.} \snm{Krafty}\corref{}\thanksref{t1,t2}\ead[label=e1]{krafty@pitt.edu}}
\and
\author[b]{\fnms{Martica} \snm{Hall}\thanksref{t2}\ead[label=e2]{hallmh@upmc.edu}}
\thankstext{t1}{Supported by NIH Grants P01-AG20677 and HL104607.}
\thankstext{t2}{Supported by NSF Grant DMS-01-04129.}
\runauthor{R. T. Krafty and M. Hall}
\affiliation{University of Pittsburgh}
\address[a]{Department of Statistics\\
University of Pittsburgh\\
2702 Cathedral of Learning\\
Pittsburgh, Pennsylvania 15260\\
USA\\
\printead{e1}}

\address[b]{Department of Psychiatry\\
University of Pittsburgh\\
3811 O'Hara Street\\
Pittsburgh, Pennsylvania 15213 \\
USA\\
\printead{e2}}

\end{aug}

\received{\smonth{12} \syear{2011}}
\revised{\smonth{9} \syear{2012}}

%
\begin{abstract}
Although many studies collect biomedical time series signals from
multiple subjects, there is a dearth of models and methods for
assessing the association between frequency domain properties of time
series and other study outcomes. This article introduces the random
Cram\'{e}r representation as a joint model for collections of time
series and static outcomes where power spectra are random functions
that are correlated with the outcomes. A canonical correlation analysis
between cepstral coefficients and static outcomes is developed to
provide a flexible yet interpretable measure of association. Estimates
of the canonical correlations and weight functions are obtained from a
canonical correlation analysis between the static outcomes and maximum
Whittle likelihood estimates of truncated cepstral coefficients. The
proposed methodology is used to analyze the association between the
spectrum of heart rate variability and measures of sleep duration and
fragmentation in a study of older adults who serve as the primary
caregiver for their ill spouse.
\end{abstract}

%

\end{frontmatter}

\section{Introduction}
\label{secintro}
Scientific and technological advances have led to an increase in the
number of studies that collect and analyze biological time series
signals from multiple subjects. In many instances, the frequency domain
properties of the time series contain interpretable physiological
information. Examples of such time series include
electroencephalographic signals [\citet{buysse2008}, \citet{qin2009wang}],
hormone concentration levels [\citet{diggle1997}, \citet
{gronfier1998}], and
heart rate variability [\citet{hall2007}, \citet{krafty2011}]. The
goal of many
such studies is to quantify the association between power spectra and
collections of correlated outcomes.

This article is motivated by a study whose objective is to better
understand the association between stress and sleep in older adults who
are the primary caregiver for their spouse. In this study, heart rate
variability and multiple measures of sleep duration and fragmentation
are collected from participants during a night of sleep. Heart rate
variability is the measure of variability in the elapsed time between
consecutive heart beats. Its power spectrum has been shown to be an
indirect measure of autonomic nervous system activity and is used by
researchers as a measure of stress [\citet{eurotask}, \citet
{hall2007}]. Measures
of sleep duration and fragmentation have been shown to be associated
with health and functioning when measured either subjectively through
self-report sleep diaries or objectively through the collection of
electrophysiological signals known as polysomnography (PSG)
[\citet{mccall1995},
\citet{hall2008},
\citet{nock2009},
\citet{silva2007},
\citet{vgontzas2010}].
We desire an analysis of these data that can illuminate the
relationship between stress and sleep by quantifying the association
between the spectrum of heart rate variability and both self-reported
and PSG derived measures of sleep.

The majority of methods for the spectral analysis of time series from
multiple subjects where spectra depend on static covariates deal
exclusively with covariates that take the form of qualitative grouping
variables [\citet{shumway1971},
\citet{diggle1997},
\citet{brillinger2001},
\citet{fokianos2008},
\citet{stoffer2010}]. These methods are not applicable when the
covariates are quantitative variables such as measures of sleep
duration and fragmentation. \citet{krafty2011} introduced the mixed
effects Cram\'{e}r representation as a model for time series data where
subject-specific power spectra depend on covariates and can account for
quantitative variables. The mixed effects Cram\'{e}r representation has
two characteristics which limit its effectiveness for modeling and
analyzing time series and correlated static outcomes. First, it assumes
a semiparametric model for log-spectra conditional on static outcomes.
As is the case in our motivating study, a semiparametric form is often
unknown and a nonparametric model is required. Second, it provides a
measure of association between time series and a static outcome
conditional on the other outcomes through a regression coefficient.
When the outcomes are correlated, extracting scientifically meaningful
information from the conditional associations provided by the multiple
regression coefficients can be challenging. In our motivating study,
clinically useful information concerning the relationship between the
spectrum of heart rate variability and the multiple correlated measures
of sleep duration and fragmentation requires parsimonious measures of
association.

To offer a nonparametric model and interpretable measures of
association between time series and sets of correlated static outcomes,
we introduce the random Cram\'{e}r representation and ensuing canonical
correlation analysis (CCA). The random Cram\'{e}r representation
considered in this article is a nonparametric joint model for time
series and sets of static outcomes where the transfer function of the
time series is random and the subject-specific log-spectra are
correlated with the static outcomes. Unlike the mixed effects Cram\'
{e}r representation of \citet{krafty2011}, no conditional semiparametric
form for the log-spectrum is assumed. The theoretical framework
introduced by \citet{eubank2008} is used to define a CCA between the
cepstral coefficients, or the Fourier coefficients of the log-spectrum
[\citet{bogert1963}], and the static outcomes. Estimates of canonical
correlations and weight functions are obtained through a procedure
which first estimates the cepstral coefficients via Whittle likelihood
regression, then performs a standard multivariate CCA between the
estimated cepstral coefficients and static outcomes.

The rest of the article is organized as follows. Section \ref
{secagewise} describes
our motivating study: the AgeWise Study. Section~\ref{secmodel}
introduces the
random Cram\'{e}r representation and CCA. The estimation procedure is
developed in Section~\ref{secest}. Section~\ref{sec5} presents the
results of a simulation
study and the proposed method is applied to data from the AgeWise Study
in Section~\ref{sec6}. A discussion is offered in Section~\ref{sec7}.

\section{The AgeWise study} \label{secagewise}
The mental and emotional stress faced by older adults who are the
primary caregiver for their ill spouse
places them at an increased risk for the development of disturbed sleep
which can effect their health and functioning [\citet{mccurry2007}]. A
goal of the AgeWise Study conducted at the University of Pittsburgh is
to gain a better understanding of the association between stress and
sleep in older adults who are the primary caregiver for their ill
spouse in order to inform the development of behavioral interventions
to enhance their sleep.

The participants in this project are $N=46$ men and women between 60--89
years of age. Each participant serves as the primary caregiver for
their spouse who is suffering from a progressive dementing illness such
as Alzheimer's or advanced Parkinson's disease. Participants were
studied during a night of in-home sleep through ambulatory PSG. The
recorded PSG signals were used to compute objective measures of total
sleep time (TST) as the number of minutes spent asleep during the
night, sleep latency (SL) as the number of minutes elapsed between
attempted sleep and sleep onset, and wakefulness after sleep onset
(WASO) as the number of minutes spent awake between sleep onset and the
final morning awakening. Upon awakening, the participants completed a
self-report sleep diary which was used to compute self-reported
measures of TST, SL, and WASO.

%
\begin{figure}

\includegraphics{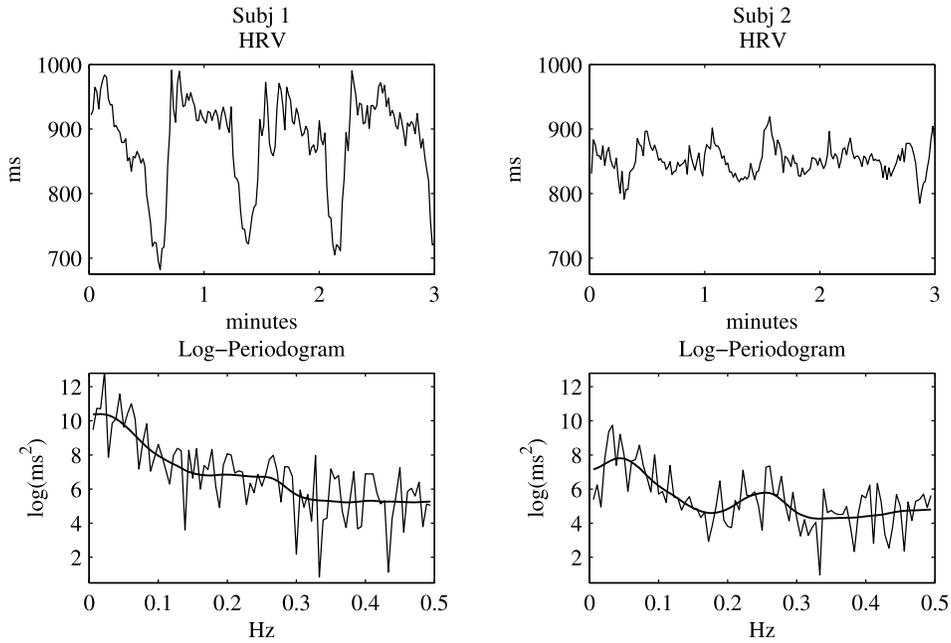}

\caption{Heart rate variability, bias-adjusted log-periodograms,
and estimated log-spectra for two subjects during the first three
minutes of stage 2 sleep. The sleep outcomes for these two subjects are
displayed in Table~\protect\ref{tabraw}.}\label{plotRawData}
\end{figure}

The ambulatory PSG included a modified 2-lead
electrode placement to collect the electrocardiogram signal
continuously throughout the night at a 512 Hz sampling rate. The
electrocardiogram signal
was digitally stored for off-line processing, cleaned of artifacts,
and used to identify the R-waves as the location of the upward\vadjust{\goodbreak}
deflection of the electrocardiogram signal associated with each
heartbeat. Interbeat intervals were then computed as the number of
milliseconds between each successive pair of R-waves to provide a
measure of the elapsed time between consecutive heart beats. Visually
scored sleep staging was temporally aligned with the interbeat
intervals for each participant to allow for the isolation of the epochs
of interbeat intervals during the first three minutes of uninterrupted
stage 2 sleep. To ensure proper physiological interpretation in
accordance with established guidelines for nonparametric spectral
analysis, we analyze the heart rate variability series generated by
sampling the cubic interpolation of the interbeat intervals versus the
R-waves at 2 Hz, resulting in time series of length $T=360$ [\citet{eurotask}].

The data for two subjects in the form of heart rate variability time
series and six measures of sleep duration and fragmentation are
displayed in the top panels of Figure~\ref{plotRawData} and in
Table~\ref{tabraw}, respectively. The primary objective of our analysis
is to illuminate the relationship between stress and sleep by obtaining
low-dimensional and interpretable measures of the association between
the spectrum of heart rate variability at the start of stage 2 sleep
and self-reported and PSG derived measures of sleep duration and fragmentation.

%
\begin{table}
\tablewidth=300pt
\caption{Time spent asleep (TST), sleep latency (SL), and wakefulness
after sleep onset (WASO) as measured by polysomnography (PSG) and
self-reported sleep diary (D) for the two subjects whose heart rate
variability are displayed in Figure~\protect\ref{plotRawData}. All sleep
outcomes are reported in minutes}
\label{tabraw}
\begin{tabular*}{300pt}{@{\extracolsep{\fill}}lcc@{}}
\hline
& \textbf{Subj. 1} & \multicolumn{1}{c@{}}{\textbf{Subj. 2}} \\
\hline
PSG-TST & 394 & 373 \\
PSG-SL & \phantom{0}11 & \phantom{0}10 \\
PSG-WASO & \phantom{0}95 & \phantom{0}78 \\
D-TST & 383 & 403 \\
D-SL & \phantom{0}20 & \phantom{00}2 \\
D-WASO & \phantom{0}45 & \phantom{0}15\\
\hline
\end{tabular*}
\end{table}

\section{Measuring association between log-spectra and static outcomes}
\label{secmodel}
\subsection{Random Cram\'{e}r representation}
This article is concerned with quantifying the association between the
spectrum of a second order stationary time series of length $T$, $
\{X_{j 1}, \ldots, X_{j T} \}$, and a $P$-dimensional vector of
correlated outcomes, $\Zj$, from $j=1,\ldots,N$ independent subjects.
In our motivating sleep study, the time series of length $T=360$ are
three minute epochs of heart rate variability from $N=46$ participants,
while $\Zj$ are $P=6$ dimensional vectors of TST, WASO, and SL as
measured by self-report sleep diary and by PSG.

The outcomes $\Zj$ are assumed to be independent and identically
distributed with $\muZ= \E(\Zj)$ and nonsingular
covariance kernel
\[
\GammaZ= \E( \Zj- \muZ) ( \Zj- \muZ)'.
\]
The time series are modeled through a random Cram\'{e}r representation
with a random mean $u_j$ and a random transfer function $\Theta_j$ that
is correlated with $\Zj$. The random transfer functions $\Theta_j$ are
independent and identically distributed complex-valued random functions
over $\mathbb{R}$ that are Hermitian, square-integrable over $[0,1]$,
and have period 1. Formally, the random Cram\'{e}r representation for
$X_{j t}$ is
\[
X_{j t} = u_j + \int_{0}^{1}
\Theta_{j}(\omega) e^{2 \pi i \omega t} \,d\Lambda_{j}(\omega),
\]
where $\Lambda_{j}$ are mutually independent identically distributed
mean-zero orthogonal increment processes over $[0,1]$ such that $\E
\llvert d\Lambda_{j} (\omega) \rrvert^2 = d
\omega$ and $\Lambda_j$ is independent of $\Theta_{j'}$, ${\bold
Z}_{j'}$, and $u_{j'}$ for all $j$ and $j'$. The time series $ \{
X_{j t}\dvtx t \in\mathbb{Z} \}$ exists with probability
one and is second order stationary.

In many applications, such as our motivating study, scientific interest
lies in the ratio of power at different frequencies. This is equivalent
to looking at linear combinations of the log-spectrum. Consequently, we
consider spectral properties on\vadjust{\goodbreak} the log-scale and define the
subject-specific log-spectrum for the $j$th subject as the random function
\[
\label{eqfj} F_j(\omega) = \log\bigl|\Theta_{j}(
\omega)\bigr|^2.
\]
To assure that the first two moments of $F_j$ exist and are bounded, it
is assumed that $\sup_{\omega\in\mathbb{R}} E \llvert\Theta
_{j}(\omega
) \rrvert^4 < \infty$ and $\inf_{\omega\in\mathbb{R}} E \llvert
\Theta_{j}(\omega) \rrvert^4 > 0$. We focus on log-spectra which possess
square integrable first derivatives and define $\mathbb{F}$ as the
space of even functions with period 1 whose first derivatives are
square integrable.

When $F_j \in\mathbb{F}$ with probability 1, the subject-specific
log-spectra possess the cosine expansion
\begin{eqnarray*}
F_{j}(\omega) &=& f_{j 0} + \sum
_{k=1}^{\infty} f_{j k} \sqrt{2} \cos(2 \pi\omega
k),
\\
f_{j 0} &=& \int_0^1
F_j(\omega) \,d \omega,
\\
f_{j k} &=& \int_0^1
F_j(\omega) \sqrt{2} \cos(2 \pi\omega k) \,d\omega,\qquad  k =1,2,\ldots,
\end{eqnarray*}
where $\mathbf{f}_j = ( f_{j k}\dvtx k =0,1,\ldots)$ is the
subject-specific cepstrum [\citet{bogert1963}]. Our analysis will explore
spectral properties of times series via the cepstral coefficients and
will make use of the covariance and cross-covariance kernels
\begin{eqnarray*}
\Gamma_f\bigl(k, k'\bigr) &=& \E\bigl[
(f_{j k} - \mu_k ) (f_{j
k'} -\mu_{k'} )
\bigr],\qquad k,k'=0,1,\ldots,
\\
\Gamma_{f Z}(k) &=& \E\bigl[ (f_{j k} - \mu_k )
( \Zj- \muZ)' \bigr],\qquad k =0,1,\ldots,
\end{eqnarray*}
where $\mu_k = E (f_{j k} )$.

\subsection{Canonical correlation analysis}\label{seccca}
To provide a parsimonious measure of association between time series
and static outcomes following a random Cram\'{e}r representation, we
will utilize the definition of CCA between two sets of second order
random variables introduced by \citet{eubank2008}. We want to find
successive linear combinations of cepstral coefficients and static
outcomes that are maximally correlated. The first canonical correlation
$\rho_1$ is defined as
\[
\rho_1^2 = \sup_{\alpha_k, \Bbeta} \cov^2
\Biggl(\sum_{k=0}^{\infty} \alpha_k
f_{j k}, \Bbeta' \Zj\Biggr)
\]
over all $\alpha_k \in\mathbb{R}$, $k=0,1,\ldots,$ and $\Bbeta\in
\mathbb{R}^P$ such that the random variables $\sum_{k=0}^{\infty}
\alpha_k f_{j k}$ and $\Bbeta' \Zj$ have unit variance. The series
$\Ba_1= (a_{1 k}\dvtx k=0,1,\ldots)$ and vector $\BB_1$ where this
maximum occurs are referred to as first canonical weights for the
cepstrum and static outcomes, while $\sum_{k=0}^{\infty}a_{1 k} f_{j
k}$ and $\BB_1'\Zj$ are first canonical variables. For\vadjust{\goodbreak} $q = 2, \ldots,
Q$ where $Q$ is the minimum of $P$ and the rank of $\Gamma_f$, the
$q$th canonical correlation $\rho_{q}$ is defined as
\[
\rho_q^2 = \sup_{\alpha_k, \Bbeta} \cov^2
\Biggl(\sum_{k=0}^{\infty} \alpha_k
f_{j k}, \Bbeta' \Zj\Biggr)
\]
over all $\alpha_k \in\mathbb{R}$, $k=0,1,\ldots,$ and $\Bbeta\in
\mathbb{R}^P$ such that $\sum_{k=0}^{\infty} \alpha_k f_{j k}$ and
$\Bbeta' \Zj$ have unit variance and are pairwise uncorrelated with
$\sum_{k=0}^{\infty} a_{q' k} f_{j k}$ and $\BB_{q'}' \Zj$ for $q' <
q$. The series $\Ba_q= (a_{q k}\dvtx k=0,1,\ldots)$ and vector
$\BB_q$ where the maximum is achieved are referred to as $q$th weight
functions for the cepstrum and static outcomes, while $\sum
_{k=0}^{\infty} a_{q k} f_{j k}$ and $\BB_q' \Zj$ are the $q$th
canonical variables. We will assume that $\GammagZ\BGamma_Z^{-1}$ is
well defined and Hilbert--Schmidt as an operator from $\mathbb{R}^P$ to
the reproducing kernel Hilbert space with reproducing kernel $\BGamma
_f$. Under this regularity condition, Theorems 1 and 2 of \citet
{eubank2008} assure the existence of the canonical correlations, weight
functions, and variables.

When $\sum_{k=0}^\infty\llvert a_{q k} \rrvert < \infty$, the $q$th
canonical variable can be represented as a linear function of the
log-spectra $F_j$,
\[
\sum_{k=0}^{\infty} a_{q k}
f_{j k} = \int_0^1 A_q(
\omega) F_j(\omega) \,d\omega,
\]
where
\[
A_q(\omega) = a_{q 0} + \sum_{k=1}^{\infty}
a_{q k} \sqrt{2} \cos(2 \pi\omega k).
\]
When it exists, we will refer to $A_q$ as the $q$th weight function for
the log-spectrum. In our application, the goal is to estimate and
analyze canonical correlations and the canonical weight functions for
the static outcomes and log-spectra. Although the canonical variables
for the cepstra do not always possess forms as integral function of
log-spectra, in Section~\ref{secfineaprox} it is shown that they can
always be approximated as such.

\subsection{Finite approximation}\label{secfineaprox}
The infinite-dimensional formulation of the CCA given above is not
conducive to real data applications. A finite-dimen\-sional approximation
can be obtained by noting that when $F_j \in\mathbb{F}$, the cepstral
coefficients decay such that $\sum_{k=1}^{\infty}k^2 f_{j k}^2 <
\infty
$ with probability 1. The decay of the cepstral coefficients was
utilized by \citet{bloomfield1973} to offer a finite-dimensional model
for the log-spectrum by truncating the cosine series at some $K < T$
such that $F_j(\omega) \approx f_{j 0} + \sum_{k=1}^{K-1} f_{j k}
\sqrt{2} \cos(2 \pi\omega k)$. Under this approximation, the
cepstrum can
be represented as the $K$-vector $\tilde{\g}_j = (f_{j 0}, \ldots,
f_{j K-1} )'$, which has $K \times K$ covariance matrix $\widetilde
{\BGamma}_f$ and $K \times P$ cross-covariance with the static outcomes
$\widetilde{\BGamma}_{f Z}$.

The CCA between the truncated cepstrum and static outcomes is a
standard multivariate CCA problem\vadjust{\goodbreak} [\citet{johnson2002}, Chapter~10.2]. To
find the canonical correlations $\widetilde{\rho}_{q}$ and variables
$\widetilde{{\mathbf a}}_{q}' \widetilde{\g}_j, \widetilde{{\mathbf
B}}_{q}' \Zj
$ between $\widetilde{\g}_j$ and $\Zj$, define $\eta_{q}$ to be the
$q$th largest eigenvalue of ${\BGamma}_Z^{-1/2}
\widetilde{\BGamma}_{f Z}' \widetilde{\BGamma}_f^{-} \widetilde
{\BGamma
}_{f Z}
{\BGamma}_Z^{-1/2}$ with associated eigenvector $\mathbf{v}_{q}$
where $\widetilde{\BGamma}_f^{-}$ is the Moore--Penrose generalized
inverse of $\widetilde{\BGamma}_f$. The
canonical correlations $\widetilde{\rho}_{q}$ and weight functions
$\widetilde{\mathbf a}_q$ and $\widetilde{\mathbf B}_q$
can be computed as
%
%
\begin{eqnarray}
\label{eqtrans} \widetilde{\rho}_{q} &=& \sqrt{\eta}_{q},
\nonumber\\
\widetilde{{\mathbf a}}_{q} &=& \widetilde{
\rho}_{q}^{-1} \widetilde{\BGamma}_f^{-}
\widetilde{\BGamma}_{f Z} \BGamma_Z^{-1/2}
\mathbf{v}_{q},
\\
\nonumber
\widetilde{\mathbf B}_q &=& \BGamma_Z^{-1/2}
\mathbf{v}_{q}.
\end{eqnarray}
These canonical correlations and weight functions are approximations of
the canonical correlations and weight functions defined in Section \ref
{seccca}. A direct consequence of Lemma 5 of \citet{eubank2008} is
that, as $K \rightarrow\infty$, $\widetilde{\rho}_q \rightarrow
\rho_q$, $ (\widetilde{\Ba}_q', 0, \ldots) \rightarrow\Ba_q$ in
$L^2(\mathbf{f}_j)$, and $\widetilde{\BB}_q \rightarrow\BB_q$.

Although the log-spectral weight function $A_q$ does not necessarily
exist, the log-spectral weight function from the finite approximation
\[
\widetilde{A}_q(\omega) = \widetilde{a}_{q 0} + \sum
_{k=1}^{K-1} \widetilde{a}_{q k}
\sqrt{2} \cos(2 \pi\omega k)
\]
is always well defined.

\section{Estimation}\label{secest}
Estimates of $\widetilde{\BGamma}_f$, $\widetilde{\BGamma}_{f Z}$, and
$\widetilde{\BGamma}_Z$ will be plugged into (\ref{eqtrans}) to obtain
estimates of the canonical correlations and weight functions. The
covariance of $\Zj$ is estimated with the standard estimator $\widehat
{\BGamma}_{Z} = (N-\break 1)^{-1} \sum_{j=1}^{N} ( \Zj- \overline{\mathbf
{Z}} ) ( \Zj- \overline{\mathbf{Z}} )'$, where $\overline
{\mathbf{Z}} = N^{-1} \sum_{j=1}^N \Zj$.

To estimate $\widetilde{\bolds\Gamma}_{f}$ and $\widetilde{\bolds
\Gamma}_{f Z}$, we consider the periodograms
\[
Y_{j \ell} = T^{-1} \Biggl\llvert\sum
_{t=1}^{T} X_{j t} e^{- 2 \pi i \ell
t/T} \Biggr
\rrvert^2,\qquad j=1,\ldots,N, \ell=1,\ldots, \biggl\lfloor\frac
{T-1}{2}
\biggr\rfloor,
\]
which are approximately independent and distributed as $e^{F_j(\ell/T)}
\chi^2_2/2$ when $T$ is large [\citet{krafty2011}, Theorem 1]. This large
sample distribution of the periodogram leads to a Whittle likelihood
[\citeauthor{whittle1953} (\citeyear{whittle1953,whittle1954})]
for truncated cepstral coefficients.
The negative log-Whittle likelihood for the truncated cepstral
coefficients of the $j$th subject is
\begin{eqnarray*}
{\cal L}_{j K} (f_0,\ldots,f_{K-1} ) &=& \sum
_{\ell=1}^{\lfloor
(T-1)/2 \rfloor}  \Biggl\{ Y_{j \ell}
e^{- [f_0 + \sum_{k=1}^{K-1}
f_{k} \sqrt{2} \cos(2 \pi k \ell/T ) ]}
\\
&&\hspace*{46pt}{} + f_0 + \sum_{k=1}^{K-1}
f_{k} \sqrt{2} \cos(2 \pi k \ell/T ) \Biggr\}.
\end{eqnarray*}
We propose using Whittle likelihood regression to estimate $\widetilde
{\BGamma}_f$ and $\widetilde{\BGamma}_{f Z}$ with
\begin{eqnarray*}
\widehat{\BGamma}_f & =& (N-1)^{-1} \sum
_{j=1}^N (\hat{\mathbf{f}}_j -
\overline{\mathbf{f}} ) (\hat{\mathbf{f}}_j -\overline{\mathbf{f}}
)' ,
\\
\widehat{\BGamma}_{f Z} & =& (N-1)^{-1} \sum
_{j=1}^N (\hat{\mathbf{ f}}_j - \overline{
\mathbf{f}} ) (\mathbf{Z}_j -\overline{\mathbf{ Z}} )',
\end{eqnarray*}
where $\hat{\mathbf{f}}_j = (\hat{f}_{j 0}, \ldots, \hat{f}_{j
K-1} )'$ minimizes ${\cal L}_{j K}$ and $\overline{\mathbf{f}} =
N^{-1} \sum_{j=1}^N \hat{\mathbf f}_j$. A Fisher's scoring algorithm for
computing $\hat{\mathbf{f}}_j$ is given in the \hyperref[app]{Appendix.}

Estimates $\widehat{\rho}_q$, $\widehat{\Ba}_q$, $\widehat{\BB
}_q$, and
$\widehat{A}_q$ of the $q$th canonical correlation and weight functions
for cepstra, static outcomes, and log-spectra are then defined for
$q=1,\ldots,Q$ as
\begin{eqnarray*}
\widehat{\rho}_{q} &=& \sqrt{\hat{\eta}_{q}} ,
\\
\widehat{\mathbf{a}}_{q} &=& \widehat{\rho}_{q}^{-1}
\widehat{\BGamma}_f^{-} \widehat{\BGamma}_{f Z}
\widehat{\BGamma}_Z^{-1/2} \widehat{ \mathbf{v}}_{q} ,
\\
\widehat{\mathbf B}_q &=& \widehat{\BGamma}_Z^{-1/2}
\widehat{\mathbf v}_{q} ,
\\
\widehat{A}_q(\omega) &=& \widehat{a}_{q 0} + \sum
_{k=1}^{K-1} \widehat{a}_{q k}
\sqrt{2} \cos(2 \pi\omega k ),\qquad \omega\in\mathbb{R},
\end{eqnarray*}
where $\widehat{\eta}_q$ is the $q$th largest eigenvalue of the $P\!\times\!P$
matrix $\widehat{\BGamma}_Z^{-1/2}
\widehat{\BGamma}_{f Z}' \widehat{\BGamma}_f^{-} \widehat{\BGamma
}_{f Z}
\widehat{\BGamma}_Z^{-1/2}$ with associated eigenvector $\widehat
{\mathbf{v}}_q$. The estimated truncated cepstral coefficients also provide
estimates of the subject-specific log-spectra as $\widehat{F}_j(\omega)
= \hat{f}_{j 0} + \sum_{k=1}^{K-1} \hat{f}_{j k} \sqrt{2} \cos(2
\pi
\omega k)$.

These estimates depend on the number of nonzero cepstral coefficients~$K$.
Simulation studies have demonstrated favorable empirical
performance of the AIC as a data driven procedure for selecting $K$ by
minimizing
\[
{\cal C}(k) = \sum_{j=1}^N {\cal
L}_{j k} (\hat{f}_{j 0},\ldots,\hat{f}_{j k-1} ) + 2 N
k.
\]

We provide Matlab code for implementing the proposed estimation
procedure in the supplemental file \citet{krafty2012B}.

\section{Simulation study}\label{sec5}
\subsection{Setting}
A simulation study was conducted to explore the empirical properties of
the proposed estimation procedure and compare it to two alternatives.
For each simulated data set, log-spectra
\[
F_j(\omega) = 5 + \sqrt{2}\cos(2 \pi\omega) + \xi_{j 0}
+ \sum_{k=1}^3 \xi_{j k} \sqrt{2}
\cos(2 \pi k \omega)
\]
were simulated where $\xi_{j k}$ are independent mean zero normal
random variables with
$\var(\xi_{j k} )=4$. Static outcomes $\Zj$ of dimension
$P=3$ were drawn as mean zero normal random vectors with covariance
matrices $\diag(4,4,4)$ such that the elements of $\Zj$ are
uncorrelated with $\xi_{j k}$, $k=0,\ldots,3$, except $\cor(\xi_{j
2}, Z_{j 1} )=0.5$ and $\cor(\xi_{j 3}, Z_{j 2} )=0.25$.
Under this setting, the canonical correlations are $\rho_1= 0.5$,
$\rho_2 = 0.25$, $\rho_3 = 0$, the weight functions for the
log-spectra are
$A_1(\omega) = \cos(4 \pi\omega)/\sqrt{2}$, $A_2(\omega) = \cos
(6 \pi
\omega)/\sqrt{2}$, and the weight functions for the static outcomes are
${\mathbf B}_1 = (0.5, 0, 0)'$, $\mathbf{B}_2 = (0, 0.5, 0)'$. After a
replicate-specific log-spectrum was simulated, its square-root was
calculated and used as the replicate-specific transfer function to
simulate the conditionally Gaussian time series $X_{j t}$ in accordance
with Theorem 2 of \citet{dai2004}. Five-hundred random samples were
drawn for each of the six combinations of $N=50,100$ and $T=30, 50,
100$. Results from additional settings under varying levels of signal
strength and smoothness are presented in the supplemental article
\citet
{krafty2012A}.

\subsection{Estimation procedures}
In addition to the proposed cepstral-based procedure, we also
investigated two alternative estimation procedures adapted from the
functional CCA literature. The CCA between two functional valued
variables has been explored by many researches including \citet
{leurgans1993}, \citeauthor{he2003} (\citeyear{he2003,he2004}), and \citet{eubank2008}. These
methods can be adapted to our setting where one set of variables,
$F_j$, is functional and observed with noise over a discrete grid
through the periodograms, and the other, $\Zj$, is multivariate.

The first alternative estimation procedure considered is an adaptation
of the algorithm for functional CCA presented in Section 6 of \citet
{eubank2008}. In this procedure, we used the penalized Whittle
likelihood of \citet{qin2009wang} to obtain smoothing spline estimators
of the subject-specific log-spectra $F_j$ with smoothing parameters
selected through direct generalized maximum likelihood. These estimated
log-spectra were discretized to form vectors of estimated log-spectra
at Fourier frequencies between 1 and $\lfloor(T-1)/2 \rfloor$; a
canonical correlation analysis was performed between these vectors and
$\Zj$. The rank of the covariance kernel of the discretized log-spectra
was selected through a cross-validation procedure which seeks to
optimize the first canonical correlation by maximizing the function
$CV_1$ discussed in Section 2.5 of \citet{he2004}.

The second alternative estimation procedure is an adaptation of the
empirical basis approach for functional CCA advocated by \citet{he2004}.
This procedure began by computing the singular value decomposition of
the sample covariance of the bias-adjusted log-periodograms, $\log
(Y_{j \ell} ) + \gamma$ where $\gamma\approx0.577$ is the
Euler--Mascheroni constant. The eigenvectors were smoothed to obtain a
functional basis. The bias-adjusted log-periodograms were then
projected onto a finite number of these basis functions and a
multivariate CCA was performed between the projections and the static
outcomes. The number of basis functions was selected through
cross-validation by maximizing $CV_1$ [\citet{he2004}, Section~2.5].

\subsection{Results}
We assessed performance through the square error of estimates of the
canonical correlations, the square error of estimated weight functions
for the static outcomes in the standard Euclidian norm, and the square
error of the vector of estimated weight functions for the log-spectra
evaluated at the Fourier frequencies in the standard Euclidian norm.
The mean and standard deviation of the square errors are displayed in
Table~\ref{tabsim}. The mean and variance of the errors of the proposed
cepstrum-based estimator were smaller than those of the two
alternatives for each parameter under every setting. The most drastic
benefit in the cepstral-based procedure was found in the estimation of
the canonical weight functions for the log-spectra. The modification of
the function CCA algorithm of \citet{eubank2008} had smaller error in
estimating the canonical correlations as compared to the empirical
basis approach, while the empirical basis approach demonstrated better
performance in estimating the weight functions of the log-spectra.

\section{Analysis of data from the AgeWise study}\label{sec6}
\subsection{Data analysis}
\begin{figure}[b]

\includegraphics{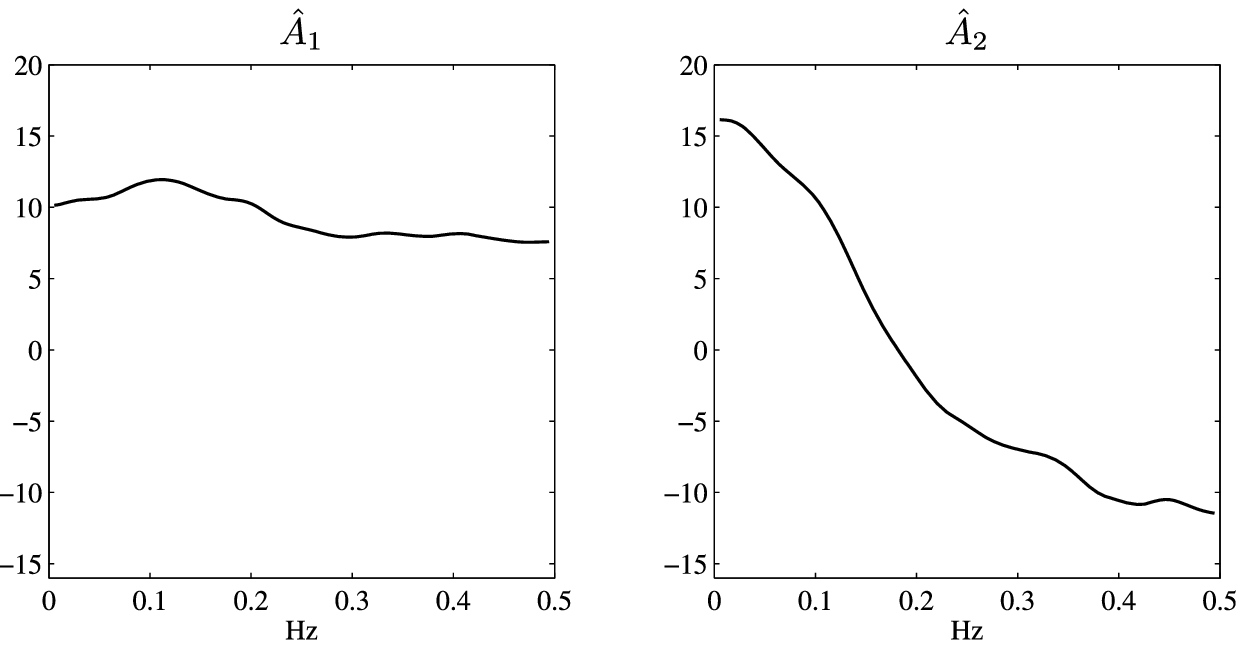}

\caption{Estimated canonical weight functions for the log-spectrum
of heart rate variability.}\label{plotWeights}
\end{figure}

We analyzed the data from the project described in Section \ref
{secagewise} that
consist of time series of heart rate variability during the first three
minutes of stage 2 sleep and $P=6$ measures of sleep duration and
fragmentation from $N=46$ participants. The mean, standard deviation,
and correlation matrix of the sleep variables are displayed in
Table~\ref{tabsleep}. To aid in the interpretation of the weight
functions, we standardized the six-dimensional vector of sleep
outcomes. The proposed procedure estimated the first two canonical
correlations as $\widehat{\rho}_1 = 0.52$ and
$\widehat{\rho}_2=0.19$; the remaining higher order correlations were
estimated to be less than 3\%. Figure~\ref{plotWeights} displays the
estimated weight functions $\widehat{A}_1, \widehat{A}_2$ of the
log-spectra, while Table~\ref{tabw} displays the estimated weight functions
$\widehat{\mathbf B}_1, \widehat{\mathbf B}_2$ of the standardized
sleep variables.

%
\begin{table}
\tabcolsep=0pt
\caption{Simulation results: The mean (standard deviation) of the
square error $\times10^{2}$
of the estimators of the first two sets of weight functions and of the
three canonical correlations. Three estimation procedures are
implemented: Cep, the proposed cepstral-based procedure; FDA,
adaptation of the functional CCA algorithm presented by \citet{eubank2008};
EB, adaptation of the empirical basis approach to functional CCA
presented by \textit{He, M{\"u}ller and Wang} (\citeyear{he2004})}
\label{tabsim}
\begin{tabular*}{\textwidth}{@{\extracolsep{\fill}}lccccccc@{}}
\hline
& \multicolumn{1}{c}{$\bolds{\widehat{A}_1}$} & \multicolumn
{1}{c}{$\bolds{\widehat{A}_2}$} &
\multicolumn{1}{c}{$\bolds{\widehat{\mathbf B}_1}$} &
\multicolumn{1}{c}{$\bolds{\widehat{\mathbf B}_2}$} &
\multicolumn{1}{c}{$\bolds{\widehat{\rho}_1}$} &
\multicolumn{1}{c}{$\bolds{\widehat{\rho}_2}$} &
\multicolumn{1}{c@{}}{$\bolds{\widehat{\rho}_3}$} \\
\hline
\multicolumn{8}{@{}l}{$N=100$, $T=100$} \\
Cep & 0.27 (0.75) & 0.79 (2.47) & 1.28 (2.19) & 3.32 (3.67) & 0.57
(0.66) &
0.85 (1.07) & 1.81 (1.62) \\
FDA & 1.87 (3.48) & 2.26 (4.37) & 1.60 (2.37) & 4.66 (4.18) & 0.80
(0.94) &
1.24 (1.58) & 1.98 (2.57) \\
EB & 1.17 (1.65) & 2.17 (2.91) & 1.73 (2.87) & 4.64 (4.15) & 1.09
(1.62) &
1.79 (1.94) & 2.87 (3.37) \\[3pt]
\multicolumn{8}{@{}l}{$N=100$, $T=50$} \\
Cep & 0.57 (1.11) & 1.67 (3.50) & 1.30 (2.08) & 3.45 (3.67) & 0.58
(0.68) &
0.84 (1.02) & 1.93 (1.68) \\
FDA & 2.03 (3.50) & 3.03 (4.86) & 1.66 (2.46) & 4.63 (4.49) & 0.80
(0.91) &
1.16 (1.44) & 1.97 (2.41) \\
EB & 1.74 (2.43) & 3.34 (4.24) & 1.61 (2.56) & 4.48 (4.24) & 1.15
(2.03) &
1.72 (1.94) & 2.79 (2.96) \\[3pt]
\multicolumn{8}{@{}l}{$N=100$, $T=30$} \\
Cep & 0.87 (0.83) & 1.61 (2.38) & 1.42 (2.21) & 3.75 (4.03) & 0.54
(0.64) &
0.84 (1.09) & 1.97 (1.67) \\
FDA & 1.99 (3.18) & 3.20 (4.52) & 1.51 (2.06) & 4.77 (4.65) & 0.80
(1.00) &
1.04 (1.19) & 2.03 (2.75) \\
EB & 1.65 (2.23) & 3.27 (4.27) & 1.62 (2.57) & 4.44 (4.34) & 1.19
(2.29) &
1.44 (1.73) & 2.54 (2.89) \\[3pt]
\multicolumn{8}{@{}l}{$N=50$, $T=100$} \\
Cep & 0.65 (2.47) & 0.98 (2.89) & 2.54 (3.32) & 5.64 (4.94) & 1.46
(1.68) &
1.85 (2.19) & 3.02 (2.74) \\
FDA & 4.26 (7.16) & 4.33 (7.41) & 3.77 (3.93) & 7.20 (5.14) & 2.36
(2.47) &
2.77 (3.07) & 3.94 (5.69) \\
EB & 1.53 (2.22) & 2.07 (2.83) & 3.44 (4.31) & 7.24 (4.88) & 2.62
(3.12) &
3.94 (3.52) & 4.82 (6.52) \\[3pt]
\multicolumn{8}{@{}l}{$N=50$, $T=50$} \\
Cep & 1.13 (2.48) & 1.93 (3.71) & 2.56 (3.38) & 5.79 (4.92) & 1.48
(1.70) &
1.93 (2.18) & 3.25 (2.89) \\
FDA & 4.63 (7.05) & 4.83 (6.75) & 3.70 (4.14) & 7.30 (5.17) & 2.55
(2.57) &
2.81 (3.20) & 4.52 (5.63) \\
EB & 2.64 (3.61) & 3.32 (4.00) & 3.60 (4.29) & 6.86 (4.81) & 2.63
(3.00) &
3.79 (3.42) & 4.95 (6.11) \\[3pt]
\multicolumn{8}{@{}l}{$N=50$, $T=30$} \\
Cep & 1.39 (1.71) & 1.84 (2.38) & 2.67 (3.33) & 5.85 (4.95) & 1.40
(1.73) &
1.99 (2.20) & 3.26 (2.94) \\
FDA & 4.52 (6.39) & 5.10 (6.65) & 3.77 (4.05) & 6.88 (5.05) & 2.55
(2.63) &
2.83 (3.41) & 4.52 (5.90) \\
EB & 3.12 (4.31) & 4.04 (5.05) & 3.50 (4.21) & 6.95 (4.92) & 2.54
(2.92) &
3.50 (3.30) & 4.55 (5.80) \\
\hline
\end{tabular*}
\end{table}

%
%
\begin{table}
\caption{Correlation matrix, means, and standard deviations of the six
sleep variables: time spent asleep (TST), sleep latency (SL), and
wakefulness after sleep onset (WASO) as measured by polysomnography
(PSG) and self-reported sleep diary (D)}\label{tabsleep}
\begin{tabular*}{\textwidth}{@{\extracolsep{\fill
}}ld{2.2}d{3.2}d{2.2}d{3.2}d{2.2}d{2.2}@{}}
\hline
& \multicolumn{1}{c}{\textbf{PSG-TST}} &
\multicolumn{1}{c}{\textbf{PSG-SL}} &
\multicolumn{1}{c}{\textbf{PSG-WASO}} &
\multicolumn{1}{c}{\textbf{D-TST}} &
\multicolumn{1}{c}{\textbf{D-SL}} &
\multicolumn{1}{c@{}}{\textbf{D-WASO}} \\
\hline
Mean (minutes) & 377.4 & 18.3 & 81.8 & 383.2 & 21.1 & 36.9 \\
Standard deviation &&&&&&\\
\quad(minutes) & 66.4 & 19.9 & 39.3 & 89.9& 31.9 & 52.8 \\
Correlation&&&&&&\\
\quad PSG-TST & 1.00& -0.12 & -0.11 & 0.45 & -0.08 & 0.17 \\
\quad PSG-SL & -0.12 & 1.00 & -0.31 & 0.06 & -0.04 & -0.22 \\
\quad PSG-WASO & -0.11 & -0.31 & 1.00 & -0.19 & 0.13 & 0.38 \\
\quad D-TST & 0.45 & 0.06 &-0.19 & 1.00 & -0.51 & -0.51 \\
\quad D-SL & -0.08 & -0.04 & 0.13 & -0.51 & 1.00 & 0.40 \\
\quad D-WASO & 0.17 & -0.22 & 0.38 & -0.51 & 0.40 & 1.00 \\
\hline
\end{tabular*}\vspace*{-3pt}
\end{table}

%

The estimated first canonical weight function for the standardized
sleep outcomes is negative for all sleep measures expect for PSG
derived WASO, which is close to zero. Note that the sum of SL, TST, and
WASO measures the total time in bed\vadjust{\goodbreak} so that large values for the first
canonical variable for the standardized sleep outcomes are associated
with less time spent in bed. The weight function for the first
canonical variable for the log-spectrum of heart rate variability is
positive at all values. Consequently, the first canonical variable for
the log-spectrum is a measure of total power or total variance.

The estimated second canonical weight function for the standardized
sleep variables is a contrast between the amount of time spent awake
during the night, as measured by both diary and PSG, and the amount of
time asleep during the night, as measured primarily by self-report
diary. The estimated second canonical weight function for the
log-spectrum is positive for frequencies less than 0.17 Hz and negative
for frequencies greater than 0.17 Hz. Recall that our analysis is on
the log-scale so that the estimated second canonical variable is a
ratio on the natural scale of power from low frequencies to power from
high frequencies. The two subjects whose data are displayed in
Figure~\ref{plotRawData} and Table~\ref{tabraw} exemplify this
association. Subject 1 displays a larger ratio of power between low and
high frequencies as compared to subject 2 with an estimated second
canonical variable of 4.13 as compared to $-0.11$. All sleep variables
for subject 1 are larger than those for subject~2 aside from diary
assessed TST. Consequently, the estimated second canonical variable for
subject 1, 0.38, is larger than that for subject 2, $-0.14$.


%
\begin{table}
\tablewidth=250pt
\caption{Estimated canonical weight functions of the standardized sleep
variables: time spent asleep (TST), sleep latency (SL), and wakefulness
after sleep onset (WASO) as measured by polysomnography (PSG) and
self-reported sleep diary (D)}
\label{tabw}
\begin{tabular*}{250pt}{@{\extracolsep{\fill}}ld{2.2}d{2.2}@{}}
\hline
& \multicolumn{1}{c}{$\bolds{\widehat{\mathbf{B}}_1}$} &
\multicolumn{1}{c@{}}{$\bolds{\widehat{\mathbf{B}}_2}$} \\
\hline
PSG-TST & -0.42 & -0.01 \\
PSG-SL & -0.52 & 0.17 \\
PSG-WASO & 0.09 & 0.33 \\
D-TST & -0.41 & -0.75 \\
D-SL & -0.55 & 0.42 \\
D-WASO & -0.51 & 0.30\\
\hline
\end{tabular*}
\end{table}

The adaptation of the functional CCA method of \citet{eubank2008} that
was explored in the simulation study was also implemented. The
subject-specific log-spectra were estimated using the smoothing spline
of \citet{qin2009wang}, while the rank of the log-spectral covariance
matrix was selected through $CV_1$ [\citet{he2004}, Section 2.5.1]. This
procedure estimated the first canonical correlation as 46\% and all
higher order canonical correlations as zero. The estimated first weight
functions for both the log-spectra and sleep outcomes were similar to
the estimates obtained through the proposed cepstral-based procedure.
However, this procedure estimated the second canonical correlation as
zero and consequently did not produce estimates of the second canonical
weight functions.

\subsection{Results}
The autonomic nervous system is classically divided into two
dynamically balanced branches: the parasympathetic branch and the
sympathetic branch. The parasympathetic branch is responsible for the
maintenance of the body at rest, while the sympathetic branch is
associated with the fight-or-flight response. Increased modulation of
the sympathetic nervous system is associated with increased power in
the spectrum of heart rate variability at low frequencies, while
increased modulation of the parasympathetic nervous system is
associated with increases in power at both low and high frequencies
[\citet{eurotask}].

The estimated first canonical variables suggest that less time in bed
is associated with increased modulation of the parasympathetic nervous
system. Excessive time spent in bed has been shown to be associated
with increased mortality and has led to the advocacy of sleep
restriction in older adults [\citet{youngstedt2004}]. The causal pathway
through which excessive time in bed is associated with mortality is
unknown and identifying possible confounders and causal intermediates
of this relationship to inform future studies is a topic of interest\vadjust{\goodbreak}
[\citet{patel2006}]. Diminished parasympathetic nervous system activity
while at rest has also been linked to mortality [\citet
{ponikowski1997}, \citet{lanza1998}]. The estimated first canonical
correlation suggests that
future studies might be able to illuminate the pathway through which
time in bed is connected to mortality by exploring the role played by
the modulation of the parasympathetic nervous system.\looseness=-1

The estimated second canonical variable for sleep is a contrast between
the time spent initiating and maintaining sleep relative to the amount
of perceived sleep and may be viewed as a measure of nocturnal
wake--sleep balance. Negative values represent less wakefulness relative
to perceived sleep; this profile is observed in healthy individuals
without clinical sleep disturbances [\citet{walsleben2004}]. In contrast,
positive wake--sleep balance values represent more wakefulness relative
to perceived sleep, as often observed in individuals with sleep
disturbances such as insomnia [\citet{carskadon1976}]. The estimated
second canonical variable for the log-spectrum is a measure of the
sympathovagal balance. Increased sympathovagal balance during sleep has
been shown to be associated with symptoms of depression and perceived
stress [\citet{hall2004}, \citet{hall2007}]. Consequently, the second
canonical
correlation suggests this simple one-dimensional measure of the
wake--sleep cycle might be useful in informing studies to develop and
evaluate behavioral therapies for improving the sleep of older adults.

\section{Discussion}\label{sec7}
This article considered an approach to analyzing the association
between the second-order spectrum of a time series and a set of static
outcomes. The random Cram\'{e}r representation provided a formal model
for these data, while the cepstrum-based CCA provided an interpretable
means of quantifying the association. This approach was motivated by
and used to analyze the association between heart rate variability
during sleep and measures of sleep duration and fragmentation in a
population of adults who are the primary caregiver for their ill
spouse. The analysis suggested a connection between stress and sleep
which can serve as a guide for designing behavioral interventions to
enhance the lives of caregivers.

The work presented in this article represents one of the first
approaches to analyzing a collection of time series whose power spectra
depend on a set of correlated outcomes and is by no means exhaustive.
This article only considered time series that are second order
stationary. Many studies which collect heart rate variability are
interested in the time-dependent spectral properties of long-term
epochs which are nonstationary [\citet{eurotask}]. A topic of future
research will be the extension of the random Cram\'{e}r representation
to the locally stationary setting through the use of a time-varying
stochastic transfer function and the development of a time dependent
cepstral coefficient-based CCA.

The CCA considered in this paper was used as a tool for exploratory
analysis. One might also be interested in inference on the canonical
correlations and weight functions. Theorem 2 of \citet{dai2004} provides
a method for simulating a time series with a given smooth spectral
density function. Another topic of future research is the development
of this sampling method to formulate a bootstrap procedure for
performing inference on the canonical correlations and weight functions.

%
\begin{appendix}\label{app}
\section*{Appendix: Fisher scoring algorithm}

The cepstral estimates $\hat{\mathbf{f}}_j$ that minimize ${\cal
L}_{j K}$
can be computed through Fisher scoring. To formulate the algorithm,
define the $K$-vectors
\begin{eqnarray*}
{\mathbf C}_{\ell} &=& \bigl\{1, \sqrt{2} \cos(2 \pi\ell/T ),
\ldots,
\sqrt{2} \cos\bigl[2 \pi\ell(K-1 )/T \bigr] \bigr\}',
\\
\mathbf{f} &=& (f_0,\ldots,f_{K-1} )'
\end{eqnarray*}
so that the negative log-Whittle likelihood for the $j$th subject can
be written as
\[
{\cal L}_{j K}(\mathbf{f}) = \sum_{\ell=1}^{\lfloor(T-1)/2 \rfloor}
\bigl(Y_{j \ell} e^{-\mathbf{C}_{\ell}' \mathbf{f} } + \mathbf
{C}_{\ell}'
{\mathbf{f}} \bigr).
\]
The algorithm is defined iteratively where the estimated cepstral
coefficients for the $j$th subject
in the $(m+1)$st iteration are
\[
\hat\mathbf{f}_j^{m+1} = \hat\mathbf{f}_j^{m}
+ \mathbf{H}^{-1} \bigl(\hat{\mathbf{f}}_j^{m}
\bigr)\mathbf{U} \bigl(\hat\mathbf{f}_j^{m} \bigr)
\]
for score function
\[
\mathbf{U} \bigl(\hat\mathbf{f}_j^{m} \bigr) =
\frac{d {\cal L}_{j
K}}{d \mathbf{f}} \Big\vert_{\mathbf{f} = \hat\mathbf{f}_j^{m}} =
\sum_{\ell
=1}^{\lfloor(T-1)/2 \rfloor}
\bigl( 1 - Y_{j \ell} e^{-\mathbf{C}_{\ell}
\hat\mathbf{f}_j^m} \bigr) \mathbf{C}_{\ell}
\]
and Fisher information matrix
\[
\mathbf{H} \bigl(\hat\mathbf{f}_j^{m} \bigr) = - \E
\biggl( \frac{d^2
{\cal L}_{j K}}{d \mathbf{f} d \mathbf{f}'} \Big\vert_{\hat
{\mathbf f}_j^{m}} \biggr) = - \sum
_{\ell= 1}^{\lfloor(T-1) / 2 \rfloor} \mathbf{C}_{\ell} {\mathbf
{C}}_{\ell}'.
\]
The algorithm continues until the change in the minimized negative
log-Whittle likelihood is below some preselected threshold. We
initialize the algorithm with the log-periodogram least squares estimators
\[
\mathbf{f}_j^0 = \bigl(\mathbf{C}'
\mathbf{C} \bigr)^{-1} \mathbf{C}' \mathbf{L}_j,
\]
where $\mathbf{L}_j = [ \log(Y_{j 1}) + \gamma, \ldots, \log(Y_{j
\lfloor(T-1)/2 \rfloor}) + \gamma]'$ and $\mathbf{C}$ is the
$\lfloor(T-1)/2\rfloor\times K$ matrix with $\ell$th row $\mathbf
{C}_{\ell}'$.
\end{appendix}

\section*{Acknowledgments}
The authors would like to thank Daniel Buysse, Anne Germain, Sati
Mazumdar, Timothy Monk, and Eric Nofzinger for their contributions to
the AgeWise study. They would also like to express their appreciation
to the editor, Tilmann Gneiting, an anonymous Associate Editor, and two
anonymous referees whose comments and suggestions greatly increased the
clarity and depth of the article.

\begin{supplement}[id=suppA]
\sname{Supplement A}
\stitle{Additional simulation results\\}
\slink[doi]{10.1214/12-AOAS601SUPPA} 
\sdatatype{.pdf}
\sfilename{aoas601\_suppa.pdf}
\sdescription{The pdf file contains the results from a more
comprehensive simulation study.}
\end{supplement}

\begin{supplement}[id=suppB]
\sname{Supplement B}
\stitle{Matlab Code}
\slink[doi]{10.1214/12-AOAS601SUPPB} 
\sdatatype{.zip}
\sfilename{aoas601\_suppb.zip}
\sdescription{The zip file contains Matlab code to run the proposed CCA
and a file demonstrating its use.}
\end{supplement}

%

\printaddresses


\begin{thebibliography}{37}

\bibitem[\protect\citeauthoryear{Bloomfield}{1973}]{bloomfield1973}
\begin{barticle}[mr]
\bauthor{\bsnm{Bloomfield},~\bfnm{P.}\binits{P.}}
(\byear{1973}).
\btitle{An exponential model for the spectrum of a scalar time series}.
\bjournal{Biometrika}
\bvolume{60}
\bpages{217--226}.
\bid{issn={0006-3444}, mr={0323048}}
\bptok{imsref}%
\end{barticle}
\endbibitem

\bibitem[\protect\citeauthoryear{Bogert, Healy and Tukey}{1963}]{bogert1963}
\begin{binproceedings}[mr]
\bauthor{\bsnm{Bogert},~\bfnm{B.~P.}\binits{B.~P.}},
  \bauthor{\bsnm{Healy},~\bfnm{M.~J.~R.}\binits{M.~J.~R.}} \AND
  \bauthor{\bsnm{Tukey},~\bfnm{J.~W}\binits{J.~W.}}
(\byear{1963}).
\btitle{The quefrency analysis of time series for echoes: Cepstrum, pseudo
  autocovariance, cross-cepstrum and saphe cracking}.
In \bbooktitle{Proceedings of the {S}ymposium on {T}ime {S}eries {A}nalysis}
(\beditor{\bfnm{M.}\binits{M.}~\bsnm{Rosenblatt}}, ed.)
\bpages{209--243}.
\bpublisher{Wiley}, \baddress{New York}.
\bptok{imsref}%
\end{binproceedings}
\endbibitem

\bibitem[\protect\citeauthoryear{Brillinger}{2001}]{brillinger2001}
\begin{bbook}[mr]
\bauthor{\bsnm{Brillinger},~\bfnm{David~R.}\binits{D.~R.}}
(\byear{2001}).
\btitle{Time Series: Data Analysis and Theory}.
\bseries{Classics in Applied Mathematics}
\bvolume{36}.
\bpublisher{SIAM},
  \blocation{Philadelphia, PA}.
\bnote{Reprint of the 1981 edition}.
\bid{doi={10.1137/1.9780898719246}, mr={1853554}}
\bptnote{check year}%
\bptok{imsref}%
\end{bbook}
\endbibitem

\bibitem[\protect\citeauthoryear{Buysse et~al.}{2008}]{buysse2008}
\begin{barticle}[author]
\bauthor{\bsnm{Buysse},~\bfnm{D.~J.}\binits{D.~J.}},
  \bauthor{\bsnm{Germain},~\bfnm{A.}\binits{A.}},
  \bauthor{\bsnm{Hall},~\bfnm{M.}\binits{M.}},
  \bauthor{\bsnm{Moul},~\bfnm{D.~E.}\binits{D.~E.}},
  \bauthor{\bsnm{Nofzinger},~\bfnm{E.~A.}\binits{E.~A.}},
  \bauthor{\bsnm{Begley},~\bfnm{A.}\binits{A.}},
  \bauthor{\bsnm{Ehler},~\bfnm{C.~L.}\binits{C.~L.}},
  \bauthor{\bsnm{Thompson},~\bfnm{W.}\binits{W.}} \AND
  \bauthor{\bsnm{Kupfer},~\bfnm{D.~J.}\binits{D.~J.}}
(\byear{2008}).
\btitle{EEG spectral analysis in primary insomnia: NREM period effects and sex
  differences}.
\bjournal{Sleep}
\bvolume{31}
\bpages{1673--1682}.
\bptok{imsref}%
\end{barticle}
\endbibitem

\bibitem[\protect\citeauthoryear{Carskadon et~al.}{1976}]{carskadon1976}
\begin{barticle}[pbm]
\bauthor{\bsnm{Carskadon},~\bfnm{M.~A.}\binits{M.~A.}},
  \bauthor{\bsnm{Dement},~\bfnm{W.~C.}\binits{W.~C.}},
  \bauthor{\bsnm{Mitler},~\bfnm{M.~M.}\binits{M.~M.}},
  \bauthor{\bsnm{Guilleminault},~\bfnm{C.}\binits{C.}},
  \bauthor{\bsnm{Zarcone},~\bfnm{V.~P.}\binits{V.~P.}} \AND
  \bauthor{\bsnm{Spiegel},~\bfnm{R.}\binits{R.}}
(\byear{1976}).
\btitle{Self-reports versus sleep laboratory findings in 122 drug-free subjects
  with complaints of chronic insomnia}.
\bjournal{Am. J. Psychiatry}
\bvolume{133}
\bpages{1382--1388}.
\bid{issn={0002-953X}, pmid={185919}}
\bptok{imsref}%
\end{barticle}
\endbibitem

\bibitem[\protect\citeauthoryear{Dai and Guo}{2004}]{dai2004}
\begin{barticle}[mr]
\bauthor{\bsnm{Dai},~\bfnm{Ming}\binits{M.}} \AND
  \bauthor{\bsnm{Guo},~\bfnm{Wensheng}\binits{W.}}
(\byear{2004}).
\btitle{Multivariate spectral analysis using {C}holesky decomposition}.
\bjournal{Biometrika}
\bvolume{91}
\bpages{629--643}.
\bid{doi={10.1093/biomet/91.3.629}, issn={0006-3444}, mr={2090627}}
\bptok{imsref}%
\end{barticle}
\endbibitem

\bibitem[\protect\citeauthoryear{Diggle and Al~Wasel}{1997}]{diggle1997}
\begin{barticle}[mr]
\bauthor{\bsnm{Diggle},~\bfnm{Peter~J.}\binits{P.~J.}} \AND
  \bauthor{\bsnm{Al~Wasel},~\bfnm{Ibrahim}\binits{I.}}
(\byear{1997}).
\btitle{Spectral analysis of replicated biomedical time series}.
\bjournal{J.~Roy. Statist. Soc. Ser. C}
\bvolume{46}
\bpages{31--71}.
\bid{doi={10.1111/1467-9876.00047}, issn={0035-9254}, mr={1452286}}
\bptnote{check related}%
\bptok{imsref}%
\end{barticle}
\endbibitem

\bibitem[\protect\citeauthoryear{Eubank and Hsing}{2008}]{eubank2008}
\begin{barticle}[mr]
\bauthor{\bsnm{Eubank},~\bfnm{R.~L.}\binits{R.~L.}} \AND
  \bauthor{\bsnm{Hsing},~\bfnm{Tailen}\binits{T.}}
(\byear{2008}).
\btitle{Canonical correlation for stochastic processes}.
\bjournal{Stochastic Process. Appl.}
\bvolume{118}
\bpages{1634--1661}.
\bid{doi={10.1016/j.spa.2007.10.006}, issn={0304-4149}, mr={2442373}}
\bptok{imsref}%
\end{barticle}
\endbibitem

\bibitem[\protect\citeauthoryear{Fokianos and Savvides}{2008}]{fokianos2008}
\begin{barticle}[mr]
\bauthor{\bsnm{Fokianos},~\bfnm{Konstantinos}\binits{K.}} \AND
  \bauthor{\bsnm{Savvides},~\bfnm{Alexios}\binits{A.}}
(\byear{2008}).
\btitle{On comparing several spectral densities}.
\bjournal{Technometrics}
\bvolume{50}
\bpages{317--331}.
\bid{doi={10.1198/004017008000000244}, issn={0040-1706}, mr={2528655}}
\bptok{imsref}%
\end{barticle}
\endbibitem

\bibitem[\protect\citeauthoryear{Gronfier and
  Brandenberger}{1998}]{gronfier1998}
\begin{barticle}[pbm]
\bauthor{\bsnm{Gronfier},~\bfnm{C.}\binits{C.}} \AND
  \bauthor{\bsnm{Brandenberger},~\bfnm{G.}\binits{G.}}
(\byear{1998}).
\btitle{Ultradian rhythms in pituitary and adrenal hormones: Their relations to
  sleep}.
\bjournal{Sleep Med. Rev.}
\bvolume{2}
\bpages{17--29}.
\bid{issn={1087-0792}, pii={S108707929890051X}, pmid={15310510}}
\bptok{imsref}%
\end{barticle}
\endbibitem

\bibitem[\protect\citeauthoryear{Hall et~al.}{2004}]{hall2004}
\begin{barticle}[author]
\bauthor{\bsnm{Hall},~\bfnm{M.}\binits{M.}},
  \bauthor{\bsnm{Vasko},~\bfnm{R.}\binits{R.}},
  \bauthor{\bsnm{Buysse},~\bfnm{D.~J.}\binits{D.~J.}},
  \bauthor{\bsnm{Ombao},~\bfnm{H.}\binits{H.}},
  \bauthor{\bsnm{Chen},~\bfnm{Q.}\binits{Q.}},
  \bauthor{\bsnm{Cashmere},~\bfnm{J.~D.}\binits{J.~D.}},
  \bauthor{\bsnm{Kupfer},~\bfnm{D.~J.}\binits{D.~J.}} \AND
  \bauthor{\bsnm{Thayer},~\bfnm{J.~F.}\binits{J.~F.}}
(\byear{2004}).
\btitle{Acute stress affects heart rate variability during sleep}.
\bjournal{Psychosomatic Medicine}
\bvolume{66}
\bpages{56--62}.
\bptok{imsref}%
\end{barticle}
\endbibitem

\bibitem[\protect\citeauthoryear{Hall et~al.}{2007}]{hall2007}
\begin{barticle}[pbm]
\bauthor{\bsnm{Hall},~\bfnm{Martica}\binits{M.}},
  \bauthor{\bsnm{Thayer},~\bfnm{Julian~F.}\binits{J.~F.}},
  \bauthor{\bsnm{Germain},~\bfnm{Anne}\binits{A.}},
  \bauthor{\bsnm{Moul},~\bfnm{Douglas}\binits{D.}},
  \bauthor{\bsnm{Vasko},~\bfnm{Raymond}\binits{R.}},
  \bauthor{\bsnm{Puhl},~\bfnm{Matthew}\binits{M.}},
  \bauthor{\bsnm{Miewald},~\bfnm{Jean}\binits{J.}} \AND
  \bauthor{\bsnm{Buysse},~\bfnm{Daniel~J.}\binits{D.~J.}}
(\byear{2007}).
\btitle{Psychological stress is associated with heightened physiological
  arousal during NREM sleep in primary insomnia}.
\bjournal{Behav. Sleep Med.}
\bvolume{5}
\bpages{178--193}.
\bid{doi={10.1080/15402000701263221}, issn={1540-2002}, pmid={17680730}}
\bptok{imsref}%
\end{barticle}
\endbibitem

\bibitem[\protect\citeauthoryear{Hall et~al.}{2008}]{hall2008}
\begin{barticle}[pbm]
\bauthor{\bsnm{Hall},~\bfnm{Martica~H.}\binits{M.~H.}},
  \bauthor{\bsnm{Muldoon},~\bfnm{Matthew~F.}\binits{M.~F.}},
  \bauthor{\bsnm{Jennings},~\bfnm{J.~Richard}\binits{J.~R.}},
  \bauthor{\bsnm{Buysse},~\bfnm{Daniel~J.}\binits{D.~J.}},
  \bauthor{\bsnm{Flory},~\bfnm{Janine~D.}\binits{J.~D.}} \AND
  \bauthor{\bsnm{Manuck},~\bfnm{Stephen~B.}\binits{S.~B.}}
(\byear{2008}).
\btitle{Self-reported sleep duration is associated with the metabolic syndrome
  in midlife adults}.
\bjournal{Sleep}
\bvolume{31}
\bpages{635--643}.
\bid{issn={0161-8105}, pmcid={2398755}, pmid={18517034}}
\bptok{imsref}%
\end{barticle}
\endbibitem

\bibitem[\protect\citeauthoryear{He, M{\"u}ller and Wang}{2003}]{he2003}
\begin{barticle}[mr]
\bauthor{\bsnm{He},~\bfnm{Guozhong}\binits{G.}},
  \bauthor{\bsnm{M{\"u}ller},~\bfnm{Hans-Georg}\binits{H.-G.}} \AND
  \bauthor{\bsnm{Wang},~\bfnm{Jane-Ling}\binits{J.-L.}}
(\byear{2003}).
\btitle{Functional canonical analysis for square integrable stochastic
  processes}.
\bjournal{J. Multivariate Anal.}
\bvolume{85}
\bpages{54--77}.
\bid{doi={10.1016/S0047-259X(02)00056-8}, issn={0047-259X}, mr={1978177}}
\bptok{imsref}%
\end{barticle}
\endbibitem

\bibitem[\protect\citeauthoryear{He, M{\"u}ller and Wang}{2004}]{he2004}
\begin{barticle}[mr]
\bauthor{\bsnm{He},~\bfnm{Guozhong}\binits{G.}},
  \bauthor{\bsnm{M{\"u}ller},~\bfnm{Hans-Georg}\binits{H.-G.}} \AND
  \bauthor{\bsnm{Wang},~\bfnm{Jane-Ling}\binits{J.-L.}}
(\byear{2004}).
\btitle{Methods of canonical analysis for functional data}.
\bjournal{J. Statist. Plann. Inference}
\bvolume{122}
\bpages{141--159}.
\bid{doi={10.1016/j.jspi.2003.06.003}, issn={0378-3758}, mr={2057919}}
\bptok{imsref}%
\end{barticle}
\endbibitem

\bibitem[\protect\citeauthoryear{Johnson and Wichern}{2002}]{johnson2002}
\begin{bbook}[author]
\bauthor{\bsnm{Johnson},~\bfnm{Richard~A.}\binits{R.~A.}} \AND
  \bauthor{\bsnm{Wichern},~\bfnm{Dean~W.}\binits{D.~W.}}
(\byear{2002}).
\btitle{Applied Multivaraite Statistical Analysis}.
\bpublisher{Prentice Hall}, \blocation{Upper Saddle River, NJ}.
\bptok{imsref}%
\end{bbook}
\endbibitem

\bibitem[\protect\citeauthoryear{Krafty, Hall and Guo}{2011}]{krafty2011}
\begin{barticle}[mr]
\bauthor{\bsnm{Krafty},~\bfnm{Robert~T.}\binits{R.~T.}},
  \bauthor{\bsnm{Hall},~\bfnm{Martica}\binits{M.}} \AND
  \bauthor{\bsnm{Guo},~\bfnm{Wensheng}\binits{W.}}
(\byear{2011}).
\btitle{Functional mixed effects spectral analysis}.
\bjournal{Biometrika}
\bvolume{98}
\bpages{583--598}.
\bid{doi={10.1093/biomet/asr032}, issn={0006-3444}, mr={2836408}}
\bptok{imsref}%
\end{barticle}
\endbibitem

\bibitem[\protect\citeauthoryear{Krafty and Hall}{2013a}]{krafty2012A}
\begin{bmisc}[author]
\bauthor{\bsnm{Krafty},~\bfnm{R.~T.}\binits{R.~T.}} \AND
  \bauthor{\bsnm{Hall},~\bfnm{M}\binits{M.}}
(\byear{2013}a).
\bhowpublished{Supplement to ``Canonical correlation analysis between time
  series and static outcomes, with application to the spectral analysis of
  heart rate variability.'' DOI:\doiurl{10.1214/12-AOAS601SUPPA}.}
\bptok{imsref}%
\end{bmisc}
\endbibitem

\bibitem[\protect\citeauthoryear{Krafty and Hall}{2013b}]{krafty2012B}
\begin{bmisc}[author]
\bauthor{\bsnm{Krafty},~\bfnm{R.~T.}\binits{R.~T.}} \AND
  \bauthor{\bsnm{Hall},~\bfnm{M}\binits{M.}}
(\byear{2013}b).
\bhowpublished{Supplement to ``Canonical correlation analysis between time
  series and static outcomes, with application to the spectral analysis of
  heart rate variability.'' DOI:\doiurl{10.1214/12-AOAS601SUPPB}.}
\bptok{imsref}%
\end{bmisc}
\endbibitem

\bibitem[\protect\citeauthoryear{Lanza et~al.}{1998}]{lanza1998}
\begin{barticle}[author]
\bauthor{\bsnm{Lanza},~\bfnm{G.}\binits{G.}},
  \bauthor{\bsnm{Guido},~\bfnm{V.}\binits{V.}},
  \bauthor{\bsnm{Galeazzi},~\bfnm{M.}\binits{M.}},
  \bauthor{\bsnm{Mustilli},~\bfnm{M.}\binits{M.}},
  \bauthor{\bsnm{Natali},~\bfnm{R.}\binits{R.}},
  \bauthor{\bsnm{Ierardi},~\bfnm{C.}\binits{C.}},
  \bauthor{\bsnm{Milici},~\bfnm{C.}\binits{C.}},
  \bauthor{\bsnm{Burzotta},~\bfnm{F.}\binits{F.}},
  \bauthor{\bsnm{Pasceri},~\bfnm{V.}\binits{V.}},
  \bauthor{\bsnm{Tomassini},~\bfnm{F.}\binits{F.}},
  \bauthor{\bsnm{Lupi},~\bfnm{A.}\binits{A.}} \AND
  \bauthor{\bsnm{Maseri},~\bfnm{A.}\binits{A.}}
(\byear{1998}).
\btitle{Prognostic role of heart rate variability in patients with recent acute
  myocardial infarction}.
\bjournal{American Journal of Caridology}
\bvolume{82}
\bpages{1323--1328}.
\bptok{imsref}%
\end{barticle}
\endbibitem

\bibitem[\protect\citeauthoryear{Leurgans, Moyeed and
  Silverman}{1993}]{leurgans1993}
\begin{barticle}[mr]
\bauthor{\bsnm{Leurgans},~\bfnm{S.~E.}\binits{S.~E.}},
  \bauthor{\bsnm{Moyeed},~\bfnm{R.~A.}\binits{R.~A.}} \AND
  \bauthor{\bsnm{Silverman},~\bfnm{B.~W.}\binits{B.~W.}}
(\byear{1993}).
\btitle{Canonical correlation analysis when the data are curves}.
\bjournal{J. Roy. Statist. Soc. Ser. B}
\bvolume{55}
\bpages{725--740}.
\bid{issn={0035-9246}, mr={1223939}}
\bptok{imsref}%
\end{barticle}
\endbibitem

\bibitem[\protect\citeauthoryear{McCall et~al.}{1995}]{mccall1995}
\begin{barticle}[pbm]
\bauthor{\bsnm{McCall},~\bfnm{W.~V.}\binits{W.~V.}},
  \bauthor{\bsnm{Turpin},~\bfnm{E.}\binits{E.}},
  \bauthor{\bsnm{Reboussin},~\bfnm{D.}\binits{D.}},
  \bauthor{\bsnm{Edinger},~\bfnm{J.~D.}\binits{J.~D.}} \AND
  \bauthor{\bsnm{Haponik},~\bfnm{E.~F.}\binits{E.~F.}}
(\byear{1995}).
\btitle{Subjective estimates of sleep differ from polysomnographic measurements
  in obstructive sleep apnea patients}.
\bjournal{Sleep}
\bvolume{18}
\bpages{646--650}.
\bid{issn={0161-8105}, pmid={8560130}}
\bptok{imsref}%
\end{barticle}
\endbibitem

\bibitem[\protect\citeauthoryear{McCurry et~al.}{2007}]{mccurry2007}
\begin{barticle}[pbm]
\bauthor{\bsnm{McCurry},~\bfnm{Susan~M.}\binits{S.~M.}},
  \bauthor{\bsnm{Logsdon},~\bfnm{Rebecca~G.}\binits{R.~G.}},
  \bauthor{\bsnm{Teri},~\bfnm{Linda}\binits{L.}} \AND
  \bauthor{\bsnm{Vitiello},~\bfnm{Michael~V.}\binits{M.~V.}}
(\byear{2007}).
\btitle{Sleep disturbances in caregivers of persons with dementia: Contributing
  factors and treatment implications}.
\bjournal{Sleep Med. Rev.}
\bvolume{11}
\bpages{143--153}.
\bid{doi={10.1016/j.smrv.2006.09.002}, issn={1087-0792}, mid={NIHMS20118},
  pii={S1087-0792(06)00106-7}, pmcid={1861844}, pmid={17287134}}
\bptok{imsref}%
\end{barticle}
\endbibitem

\bibitem[\protect\citeauthoryear{Nock et~al.}{2009}]{nock2009}
\begin{barticle}[author]
\bauthor{\bsnm{Nock},~\bfnm{Nora~L}\binits{N.~L.}},
  \bauthor{\bsnm{Li},~\bfnm{Li}\binits{L.}},
  \bauthor{\bsnm{Larkin},~\bfnm{Emma~K}\binits{E.~K.}},
  \bauthor{\bsnm{Patel},~\bfnm{Sanjay~R}\binits{S.~R.}} \AND
  \bauthor{\bsnm{Redline},~\bfnm{Susan}\binits{S.}}
(\byear{2009}).
\btitle{Empirical evidence for ``syndrome Z'': A hierarchical 5-factor model of
  the metabolic syndrome incorporating sleep disturbance measures.}
\bjournal{Sleep}
\bvolume{32}
\bpages{615--622}.
\bptok{imsref}%
\end{barticle}
\endbibitem

\bibitem[\protect\citeauthoryear{Patel et~al.}{2006}]{patel2006}
\begin{barticle}[author]
\bauthor{\bsnm{Patel},~\bfnm{Sanjay}\binits{S.}},
  \bauthor{\bsnm{Malhotra},~\bfnm{Atul}\binits{A.}},
  \bauthor{\bsnm{Gottlieb},~\bfnm{Daniel~J.}\binits{D.~J.}},
  \bauthor{\bsnm{White},~\bfnm{David~P.}\binits{D.~P.}} \AND
  \bauthor{\bsnm{Hu},~\bfnm{Frank~B.}\binits{F.~B.}}
(\byear{2006}).
\btitle{Correlates of long sleep duration}.
\bjournal{Sleep}
\bvolume{29}
\bpages{881--889}.
\bptok{imsref}%
\end{barticle}
\endbibitem

\bibitem[\protect\citeauthoryear{Ponikowski et~al.}{1997}]{ponikowski1997}
\begin{barticle}[pbm]
\bauthor{\bsnm{Ponikowski},~\bfnm{P.}\binits{P.}},
  \bauthor{\bsnm{Anker},~\bfnm{S.~D.}\binits{S.~D.}},
  \bauthor{\bsnm{Chua},~\bfnm{T.~P.}\binits{T.~P.}},
  \bauthor{\bsnm{Szelemej},~\bfnm{R.}\binits{R.}},
  \bauthor{\bsnm{Piepoli},~\bfnm{M.}\binits{M.}},
  \bauthor{\bsnm{Adamopoulos},~\bfnm{S.}\binits{S.}},
  \bauthor{\bsnm{Webb-Peploe},~\bfnm{K.}\binits{K.}},
  \bauthor{\bsnm{Harrington},~\bfnm{D.}\binits{D.}},
  \bauthor{\bsnm{Banasiak},~\bfnm{W.}\binits{W.}},
  \bauthor{\bsnm{Wrabec},~\bfnm{K.}\binits{K.}} \AND
  \bauthor{\bsnm{Coats},~\bfnm{A.~J.}\binits{A.~J.}}
(\byear{1997}).
\btitle{Depressed heart rate variability as an independent predictor of death
  in chronic congestive heart failure secondary to ischemic or idiopathic
  dilated cardiomyopathy}.
\bjournal{Am. J. Cardiol.}
\bvolume{79}
\bpages{1645--1650}.
\bid{issn={0002-9149}, pii={S0002914997002154}, pmid={9202356}}
\bptok{imsref}%
\end{barticle}
\endbibitem

\bibitem[\protect\citeauthoryear{Qin and Wang}{2008}]{qin2009wang}
\begin{barticle}[mr]
\bauthor{\bsnm{Qin},~\bfnm{Li}\binits{L.}} \AND
  \bauthor{\bsnm{Wang},~\bfnm{Yuedong}\binits{Y.}}
(\byear{2008}).
\btitle{Nonparametric spectral analysis with applications to seizure
  characterization using {EEG} time series}.
\bjournal{Ann. Appl. Stat.}
\bvolume{2}
\bpages{1432--1451}.
\bid{doi={10.1214/08-AOAS185}, issn={1932-6157}, mr={2655666}}
\bptnote{check year}%
\bptok{imsref}%
\end{barticle}
\endbibitem

\bibitem[\protect\citeauthoryear{Shumway}{1971}]{shumway1971}
\begin{barticle}[author]
\bauthor{\bsnm{Shumway},~\bfnm{R.~H.}\binits{R.~H.}}
(\byear{1971}).
\btitle{On detecting a signal in $N$ stationarily correlated noise series}.
\bjournal{Technometrics}
\bvolume{13}
\bpages{499--519}.
\bptok{imsref}%
\end{barticle}
\endbibitem

\bibitem[\protect\citeauthoryear{Silva et~al.}{2007}]{silva2007}
\begin{barticle}[pbm]
\bauthor{\bsnm{Silva},~\bfnm{Graciela~E.}\binits{G.~E.}},
  \bauthor{\bsnm{Goodwin},~\bfnm{James~L.}\binits{J.~L.}},
  \bauthor{\bsnm{Sherrill},~\bfnm{Duane~L.}\binits{D.~L.}},
  \bauthor{\bsnm{Arnold},~\bfnm{Jean~L.}\binits{J.~L.}},
  \bauthor{\bsnm{Bootzin},~\bfnm{Richard~R.}\binits{R.~R.}},
  \bauthor{\bsnm{Smith},~\bfnm{Terry}\binits{T.}},
  \bauthor{\bsnm{Walsleben},~\bfnm{Joyce~A.}\binits{J.~A.}},
  \bauthor{\bsnm{Baldwin},~\bfnm{Carol~M.}\binits{C.~M.}} \AND
  \bauthor{\bsnm{Quan},~\bfnm{Stuart~F.}\binits{S.~F.}}
(\byear{2007}).
\btitle{Relationship between reported and measured sleep times: The sleep heart
  health study (SHHS)}.
\bjournal{J.~Clin. Sleep Med.}
\bvolume{3}
\bpages{622--630}.
\bid{issn={1550-9389}, pmcid={2045712}, pmid={17993045}}
\bptok{imsref}%
\end{barticle}
\endbibitem

\bibitem[\protect\citeauthoryear{Stoffer et~al.}{2010}]{stoffer2010}
\begin{barticle}[mr]
\bauthor{\bsnm{Stoffer},~\bfnm{David~S.}\binits{D.~S.}},
  \bauthor{\bsnm{Han},~\bfnm{Sangdae}\binits{S.}},
  \bauthor{\bsnm{Qin},~\bfnm{Li}\binits{L.}} \AND
  \bauthor{\bsnm{Guo},~\bfnm{Wensheng}\binits{W.}}
(\byear{2010}).
\btitle{Smoothing spline {ANOPOW}}.
\bjournal{J.~Statist. Plann. Inference}
\bvolume{140}
\bpages{3789--3796}.
\bid{doi={10.1016/j.jspi.2010.04.043}, issn={0378-3758}, mr={2674166}}
\bptok{imsref}%
\end{barticle}
\endbibitem

\bibitem[\protect\citeauthoryear{Task Force of the ESC/ASPE}{1996}]{eurotask}
\begin{bmisc}[author]
\borganization{Task Force of the ESC/ASPE}
(\byear{1996}).
\bhowpublished{Heart rate variability---standards of measurement, physiological
  interpretation, and clinical use. \textit{Circulation} \textbf{93}
  1043--1065.}
\bptok{imsref}%
\end{bmisc}
\endbibitem

\bibitem[\protect\citeauthoryear{Vgontzas et~al.}{2010}]{vgontzas2010}
\begin{barticle}[pbm]
\bauthor{\bsnm{Vgontzas},~\bfnm{Alexandros~N.}\binits{A.~N.}},
  \bauthor{\bsnm{Liao},~\bfnm{Duanping}\binits{D.}},
  \bauthor{\bsnm{Pejovic},~\bfnm{Slobodanka}\binits{S.}},
  \bauthor{\bsnm{Calhoun},~\bfnm{Susan}\binits{S.}},
  \bauthor{\bsnm{Karataraki},~\bfnm{Maria}\binits{M.}},
  \bauthor{\bsnm{Basta},~\bfnm{Maria}\binits{M.}},
  \bauthor{\bsnm{Fern{\'{a}}ndez-Mendoza},~\bfnm{Julio}\binits{J.}} \AND
  \bauthor{\bsnm{Bixler},~\bfnm{Edward~O.}\binits{E.~O.}}
(\byear{2010}).
\btitle{Insomnia with short sleep duration and mortality: The penn state
  cohort}.
\bjournal{Sleep}
\bvolume{33}
\bpages{1159--1164}.
\bid{issn={0161-8105}, pmcid={2938855}, pmid={20857861}}
\bptok{imsref}%
\end{barticle}
\endbibitem

\bibitem[\protect\citeauthoryear{Walsleben et~al.}{2004}]{walsleben2004}
\begin{barticle}[pbm]
\bauthor{\bsnm{Walsleben},~\bfnm{Joyce~A.}\binits{J.~A.}},
  \bauthor{\bsnm{Kapur},~\bfnm{Vishesh~K.}\binits{V.~K.}},
  \bauthor{\bsnm{Newman},~\bfnm{Anne~B.}\binits{A.~B.}},
  \bauthor{\bsnm{Shahar},~\bfnm{Eyal}\binits{E.}},
  \bauthor{\bsnm{Bootzin},~\bfnm{Richard~R.}\binits{R.~R.}},
  \bauthor{\bsnm{Rosenberg},~\bfnm{Carl~E.}\binits{C.~E.}},
  \bauthor{\bsnm{O'Connor},~\bfnm{George}\binits{G.}} \AND
  \bauthor{\bsnm{Nieto},~\bfnm{F.~Javier}\binits{F.~J.}}
(\byear{2004}).
\btitle{Sleep and reported daytime sleepiness in normal subjects: The sleep
  heart health study}.
\bjournal{Sleep}
\bvolume{27}
\bpages{293--298}.
\bid{issn={0161-8105}, pmid={15124725}}
\bptok{imsref}%
\end{barticle}
\endbibitem

\bibitem[\protect\citeauthoryear{Whittle}{1953}]{whittle1953}
\begin{barticle}[mr]
\bauthor{\bsnm{Whittle},~\bfnm{P.}\binits{P.}}
(\byear{1953}).
\btitle{Estimation and information in stationary time series}.
\bjournal{Ark. Mat.}
\bvolume{2}
\bpages{423--434}.
\bid{issn={0004-2080}, mr={0060797}}
\bptok{imsref}%
\end{barticle}
\endbibitem

\bibitem[\protect\citeauthoryear{Whittle}{1954}]{whittle1954}
\begin{barticle}[author]
\bauthor{\bsnm{Whittle},~\bfnm{P.}\binits{P.}}
(\byear{1954}).
\btitle{Some recent contributions to the theory of stationary processes}.
\bjournal{A~Study in the Analysis of Stationary Time Series}
\bvolume{2}
\bpages{196--228}.
\bptok{imsref}%
\end{barticle}
\endbibitem

\bibitem[\protect\citeauthoryear{Youngstedt and Kripke}{2004}]{youngstedt2004}
\begin{barticle}[author]
\bauthor{\bsnm{Youngstedt},~\bfnm{S.~D.}\binits{S.~D.}} \AND
  \bauthor{\bsnm{Kripke},~\bfnm{D.~F.}\binits{D.~F.}}
(\byear{2004}).
\btitle{Long sleep and mortality: Rationale for sleep restriction}.
\bjournal{Sleep Med. Rev.}
\bvolume{8}
\bpages{159--174}.
\bptok{imsref}%
\end{barticle}
\endbibitem

\end{thebibliography}
\end{document}